\documentclass[onecolumn,showpacs,pra,aps]{revtex4}
\usepackage{dcolumn}
\usepackage{bm}
\usepackage{graphicx}
\usepackage{mathrsfs}
\begin{document}
\title{Two-Photon Beatings Using Biphotons Generated from a Two-Level System}
\author{Jianming Wen,$^1$\footnote{jianm1@umbc.edu} Shengwang Du,$^2$ Morton H.
Rubin,$^1$ and Eun Oh$^{3,4}$} \address{$^1$Physics Department,
University of Maryland, Baltimore County, Baltimore, Maryland 21250, USA\\
$^2$Department of Physics, The Hong Kong University of Science and Technology, Clear
Water Bay, Kowloon, Hong Kong, China\\
$^3$Remote Sensing Division, U.S. Naval Research Laboratory, Washington, D.C. 20375, USA\\
$^4$Physics Department, University of Virginia, Charlottesville, Virginia 22904, USA}

\date{\today}

\begin{abstract}
We propose a two-photon beating experiment based upon biphotons generated from a resonant pumping two-level system operating in a backward geometry. On the one hand, the linear optical-response leads biphotons produced from two sidebands in the Mollow triplet to propagate with tunable refractive indices, while the central-component propagates with unity refractive index. The relative phase difference due to different refractive indices is analogous to the pathway-length difference between long-long and short-short in the original Franson interferometer. By subtracting the linear Rayleigh scattering of the pump, the visibility in the center part of the two-photon beating interference can be ideally manipulated among [0, 100$\%$] by varying the pump power, the material length, and the atomic density, which indicates a Bell-type inequality violation. On the other hand, the proposed experiment may be an interesting way of probing the quantum nature of the detection process. The interference will disappear when the separation of the Mollow peaks approaches the fundamental timescales for photon absorption in the detector.
\end{abstract}

\pacs{42.50.Xa, 03.65.Ta, 03.65.Ud, 42.50.-p}

\maketitle

\section{Introduction}
Entangled paired photons or biphotons provide an unprecedented tools for research in fundamental physics such as quantum information processing \cite{QIP} and tests of fundamentals of quantum mechanics \cite{bell}. Biphotons produced from either spontaneous parametric down conversion (SPDC) \cite{SPDC} or atomic cascade transitions have already been used in optical experimental tests of fundamentals of quantum theory, and have demonstrated violations of Bell's inequalities \cite{aspect,shih}. Most of experimental tests of the inequalities have involved the polarization entanglement. Rarity and Tapster \cite{rarity} have reported experiments with momentum entanglement of the beams. The
experiment proposed by Franson \cite{franson} concerning a Bell inequality for nonpolarization variables, relies on the entanglement of a continuous variable, energy. There are number of experimental realizations \cite{experiment,kwiat} of the Franson interferometer. Kwiat \textit{et al}. \cite{kwiat} have achieved sinusoidal fringes of the two-photon interference with visibility greater than $70.7\%$. Such high visibility is required to confirm violation of a Bell-type inequality. In this paper, we propose a new type of two-photon beating experiment by using biphotons
generated from a two-level system operating in the resonant-pumping case \cite{osa07}.

Recently, the coincidence counting rate of paired photons generated from a two-level atomic system exhibits a damped Rabi oscillation and photon anti-bunching-like effect \cite{wen,du}. These features are caused by the destructive interference between two types of biphoton generation [or four-wave mixings (FWMs)] -- sideband-sideband and central-central in the \textit{Mollow triplet}. The two-photon state is therefore a coherent superposition of these two FWMs. We have also emphasized that the coupling coefficients in the coupled field operators equations for paired-photon generation are different from those in the classical FWM equations in a two-level system \cite{wen,du}. The two-photon wavepacket created in such a two-level system is generally a convolution of the phase matching and the third-order nonlinear susceptibility, as in the three-level \cite{wen2} and four-level \cite{wen21,wen3,du2008} double-$\Lambda$ systems. A further analysis shows that these two kinds of biphotons created in a two-level system are associated with different refractive indices \cite{quang}. The linear optical response of the paired photons from two sidebands in the Mollow triplet \cite{mollow} implies that these biphotons propagate with tunable refractive indices and vanishing absorption, while biphotons from the central component propagate with unity refraction of index. These different refractive indices induce a relative phase difference between the twin-photon propagation through the two-level medium, and play the same role as the pathway-length difference between the short-short (S-S) and long-long (L-L) in the original Franson's experiment. It should be noted that two-photon interference disappears, if the detection system can spectrally distinguish these two FWM processes. Such a filtering would destroy the biphoton generated by the sidebands (or central components). On the other hand, the proposed experiment may be an interesting way of probing the quantum nature of the detection process. The interference should disappear when the separation of the Mollow peaks approaches the fundamental timescales for photon absorption in the detector.

The proposed two-photon beating experiment depends on the interference between the probability amplitudes of two types of biphotons emitted at various time from a two-level system. In the counter-propagating geometry, the two-photon interference has a symmetrically damped Rabi oscillation with the oscillation period inversely proportional to the effective Rabi frequency. Moreover, due to the relative phase difference caused by the refractive indices of two types of biphotons, the visibility of the center part of the two-photon coincidence pattern can be altered from 0 to $100\%$ in the ideal case, by subtracting the linear Rayleigh scattering of the input pump beam. This change indicates a switch between photon anti-bunching-like and photon bunching-like effects. Alternatively, the interference between two types of FWMs can be switched from destructive to constructive by changing the pump Rabi frequency (or the pump power), the length of the medium, and the atomic density. Furthermore, because the proposed beating experiment has direct similarity with the Franson interferometer, such a high visibility (ideally, 100$\%$) of the center part in the two-photon beating pattern, greater than 70.7$\%$ as predicted by quantum mechanics, implies a violation of Bell's inequality.

The paper is organized as follows. In Sec.~II, we describe the linear and nonlinear optical responses for different photon pairs generated in the Mollow triplet. In Sec.~III, we present the detailed analysis of the two-photon beatings using biphotons created from a two-level system. We emphasize that the two-photon state is a coherent superposition of two FWM processes, and paired photons produced from these processes experience different phases. The analog between the proposed experiment and the Franson interferometer is built by analyzing the two-photon state. Finally, we give our conclusions in Sec.~IV.

\section{Linear and nonlinear optical responses}
Consider a medium of identical two-level \textit{atoms} (or \textit{molecules}) initially prepared in their ground level $|g\rangle$. We configure our system with presence of a retro-reflected pump beam. In this configuration, the generated paired photons are spontaneously emitted in opposite directions and are detected by two photodetectors D$_1$ and D$_2$, as illustrated in Fig.~1(a). The strong pump field with angular frequency $\omega_p$ and Rabi frequency $\Omega_p$ is applied to the quantum transition $|g\rangle\rightarrow|e\rangle$ with a detuning $\Delta=\omega_p-\omega_{eg}$, where $\omega_{eg}$ is the atomic transition frequency. The atomic oscillators are driven into the nonlinear regime where the absorbed two counter-propagating pump photons are re-radiated at frequencies $\omega_3=\omega_p+\delta$ and $\omega_4=\omega_p-\delta$ as seen in Fig.~1(b). It is found that the linear optical response to the $\omega_3$ field and the third-order nonlinear susceptibility of the $\omega_3$ field are \cite{wen}
\begin{eqnarray}
\chi_3(\delta)=\frac{N|\mu|^2/\hbar\epsilon_0}{\Delta+\delta+i\gamma_2}\bigg[1-\frac{|\Omega_p|^2(\delta+2i\gamma_2)
(\delta-\Delta+i\gamma_2)}{2D(\delta)(\Delta-i\gamma_2)}\bigg],\label{eq:linear}
\end{eqnarray}
\begin{eqnarray}
\chi^{(3)}_3(\delta)&=&\frac{2N|\mu|^4(\delta+i\gamma_2)}{\hbar^3\epsilon_0D(\delta)(\Delta+i\gamma_2)}\nonumber\\
&\approx&\frac{N|\mu|^4}{\hbar^3\epsilon_0(\Delta+i\gamma_2)}\Bigg[\frac{1}{(\delta+i\Gamma_0)(\delta+\Omega_e+i\Gamma)}
+\frac{1}{(\delta+i\Gamma_0)(\delta-\Omega_e+i\Gamma)}\Bigg],
\label{eq:nonlinearity}
\end{eqnarray}
with
\begin{eqnarray}
D(\delta)=(\delta+i\gamma_g)(\delta+\Delta+i\gamma_2)(\delta-\Delta+i\gamma_2)-|\Omega_p|^2(\delta+i\gamma_2). \label{eq:D}
\end{eqnarray}
In Eqs.~(\ref{eq:linear})-(\ref{eq:D}), $\gamma_g$ is the decay rate of the ground state, $\gamma_2$ is the dephasing rate, $\Gamma_0$ and $\Gamma$ are the linewidths of biphotons produced from the central component and two sidebands in the Mollow triplet, $N$ is the atomic density, and $\mu$ is the dipole matrix element, respectively. In the resonant pumping case, $\Gamma_0$ reduces to $\gamma_2$ and $\Gamma$ becomes $(\gamma_g+\gamma_2)/2$. To emphasize two FWM processes occurring in the system, we have decomposed $\chi^{(3)}_3(\delta)$ into two terms as shown in the second line of Eq.~(\ref{eq:nonlinearity}). The Mollow triplet resonances embedded in $D(\delta)$ is depicted by the third-order nonlinear susceptibility $|\chi^{(3)}(\delta)|$ shown in Fig.~1(c).

To look at the optical response of the medium, we begin with Eqs.~(\ref{eq:linear}) and (\ref{eq:nonlinearity}). It is imperative to examine the linear optical response of paired photons generated from the central or two sidebands in the Mollow triplet and ascertain their propagation properties. In Fig.~1(d), we plot real (red) and imaginary (blue) parts of the linear susceptibility of the $\omega_3$ field [Eq.~(\ref{eq:linear})] for the resonant pumping case. The slopes of the real part of $\chi_3(\delta)$ directly govern the group velocities of paired photons crossing the medium while the imaginary part dictates types of spectral transmission windows.

As shown in Fig.~1(d), biphotons originating from the central component experience an anomalous dispersion inside the medium. Consequently, photons generated from the central component undergo a superluminal propagation (group velocity greater than $c$, the speed of light in vacuum). This can be understood from the gain-assisted superluminal light propagation mechanism \cite{wang} where the anomalous dispersion plays the dominant role. Also shown in Fig.~1(d), near the region of the sideband frequencies $\delta=\omega_p\pm|\Omega_p|$ the indices of refraction reach a maximum (or minimum) value and the absorption of the medium vanishes. In other words, paired photons originating from the two sidebands propagate with tunable refractive indices accompanied by vanishing absorption (of the medium). For the photons originating from the central component, they propagate with unity index of refraction accompanied by (strong) absorption. This difference in refractive indices between the two types of biphotons generated from either central or sidebands serves as the pathway-length difference in the Franson's experiment.

\section{Two-photon beatings}
In the case of spontaneous emission, the two-photon amplitude at the two photodetectors can in general be represented as
\begin{eqnarray}
A&=&\langle0|E^{(+)}_1(\tau_1)E^{(+)}_2(\tau_2)|\Psi\rangle\nonumber\\
&=&\frac{1}{\sqrt{2}}\Big(e^{-\gamma_g\tau}e^{i\phi}-e^{-\gamma_2\tau}e^{i\Omega_e\tau}\Big),\label{eq:state}
\end{eqnarray}
where $\tau=|\tau_1-\tau_2|$ is the timing difference between two detectors, $\Omega_e=\sqrt{\Delta^2+|\Omega_p|^2}$ is the effective Rabi frequency (in the resonant pumping case as discussed in this paper, it is only determined by the pump Rabi frequency), and $E^{(+)}_j(\tau_j)$ is the annihilation operator of the free-space propagating electric field at detector D$_j$ ($j=1,2$), respectively. In the second line of Eq.~(\ref{eq:state}), the first term in the bracket implies that biphotons are produced from the central component $|1_0,1_0\rangle$ (which have the same frequency as the pump), and the second term stands for biphotons generated from two correlated sidebands $|1_{\Omega_e},1_{-\Omega_e}\rangle$ (whose central frequencies are peaked away from the pump frequency by $\Omega_e$ and $-\Omega_e$), respectively. The spectral separation between these two FWM processes is equal to $\Omega_e$. The two-photon amplitude shown in Eq.~(\ref{eq:state}) is a coherent superposition of two FWM processes. The ``-" sign in Eq.~(\ref{eq:state}) indicates a $\pi$ phase shift between two FWM processes. The phase term $e^{i\Omega_e\tau}$ arises from the Rabi oscillation due to the dynamic Stark shift of the atomic levels. $\phi$ is the relative phase difference between two types of biphotons generated as they pass through the medium with length $L$,
\begin{eqnarray}
\phi=\frac{[2k_0-(k_{\Omega_e}+k_{-\Omega_e})]L}{2},\label{eq:phi}
\end{eqnarray}
where $k_{\Omega_e}$ and $k_{-\Omega_e}$ are central wave numbers of two sidebands inside the material and $k_0$ the central wave number of the central component. The dispersion within the medium is $k_j=n_j\omega_j/c$, and the refractive index $n_j=\sqrt{1+\chi'_j}$ where $\chi'$ represents the real part of the linear susceptibility $\chi$. In the resonant pumping case, $\chi'$ at two sideband frequencies is opposite in sign: positive at $\omega_p+|\Omega_p|$ but negative at $\omega_p-|\Omega_p|$ [Fig.~1(d)]. This property sets a useful region of the refractive index for two sidebands, i.e., $n_{-\Omega_e}\in(0,1)$ and $n_{\Omega_e}\in(2,1)$. We note that $n_{\Omega_e}$ can be enhanced without absorption \cite{scully}. If $n_{\Omega_e}\geq2$, however, $n_{-\Omega_e}$ becomes imaginary, indicating an
evanescent exponentially decreasing field and destroying the efficiency of biphoton generation. Eq.~(\ref{eq:state}) also shows that the two types of biphotons have different spectral widths $\gamma_g$ and $\gamma_2$. $\Omega_e\tau$ and $\phi$ together determine the pattern of the two-photon temporal correlation, as shown below. To compare our result with the Franson interferometer, we know that the two-photon amplitude presented in Eq.~(4) can also be visualized as a path-entangled biphoton wavepacket, as paired photons from central component in the Mollow triplet traverse the short-short pathways while biphotons from correlated sidebands traverse the long-long pathways to cross over the medium. Even though the amplitude given in Eq.~(\ref{eq:state}) is physically different from that required in the Franson experiment, the similarity between these two states offers another way to look at the path-entangled problem. One consequence is to examine the violation of Bell-type inequality by studying the two-photon beatings.
%%%%%%%%%%%%%%%%%%%%%%%%%%%%%%%%%%%%%%%%%%%%%%%%%%
\begin{figure}[htb]
\centering\includegraphics[scale=0.8]{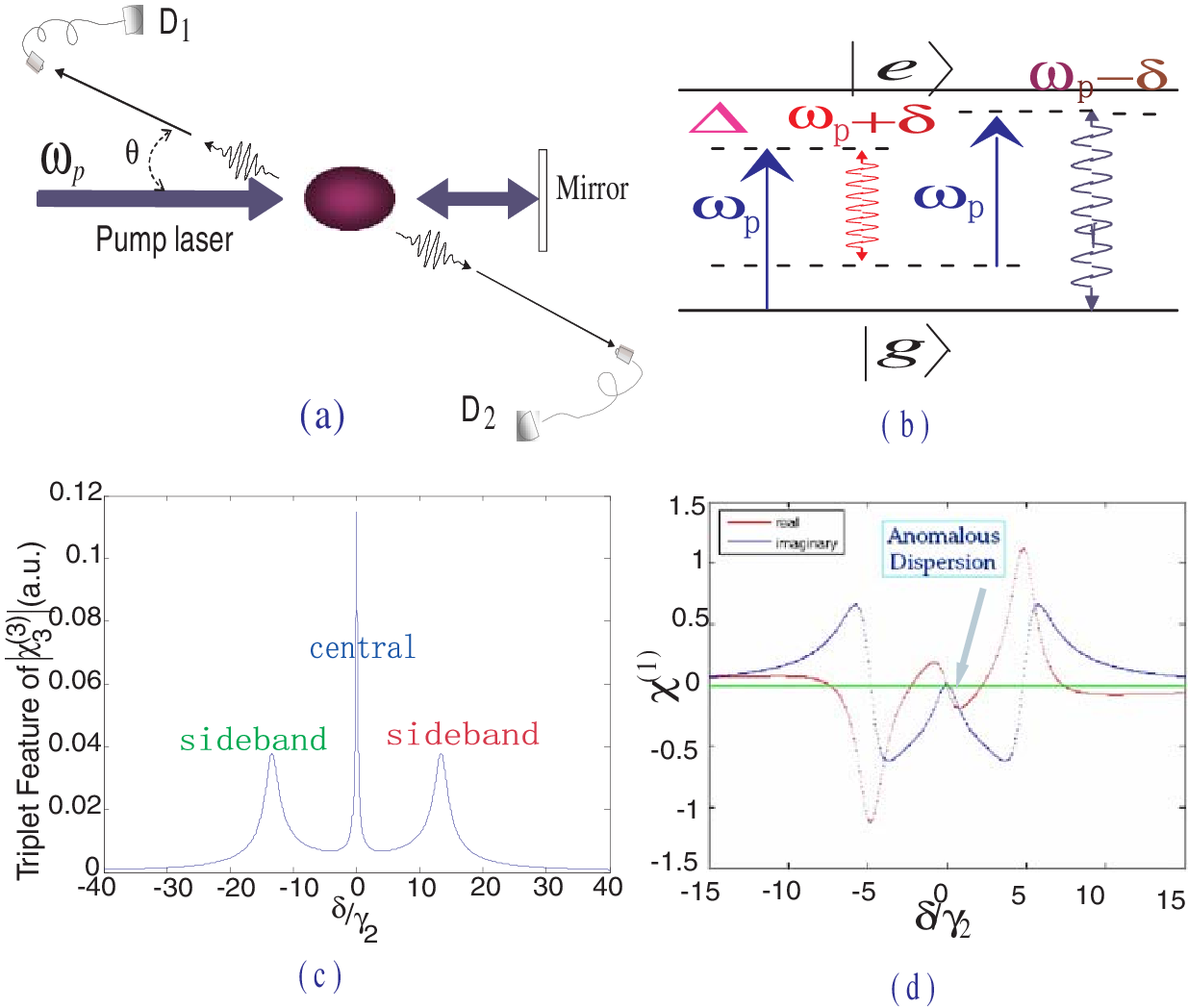}
\caption{(Color Online) (a) Two-photon coincidence measurements implemented in the counter-propagation geometry. (b) In the presence of a retro-reflected, strong pump beam at angular frequency $\omega_p$, weak fields at frequencies $\omega_p+\delta$ and $\omega_p-\delta$ are generated, shown by wavy arrows. (c) Spectrum of the third-order nonlinear susceptibility $|\chi^{(3)}_3(\delta)|$ shows a feature of the Mollow triplet. (d) Linear optical response in the resonant pumping case: the red line is the real part of $\chi_3(\delta)$ and the blue one is the imaginary part. The central component will experience anomalous dispersion.} \label{Fig.1}
\end{figure}
%%%%%%%%%%%%%%%%%%%%%%%%%%%%%%%%%%%%%%%%%%%%%%%%%%

%Eq.~(\ref{eq:state}) can be also visualized as an
%intuitive picture of the two-photon \textit{dip} experiment
%\cite{HOM}, where the probability amplitudes of all the
%indistinguishable processes should be summed together. In the
%resonant pumping case, $|1_0,1_0\rangle$ in Eq.~(\ref{eq:state})
%can be considered as the state of paired photons transmitted after
%the beam splitter while $|1_{\Omega_e},1_{-\Omega_e}\rangle$ can
%be modeled as the state of twin photons reflected before the detection in the
%\textit{dip} experiment. %The analogy with the two-photon dip experiment is
%schematically shown in Fig.~1(d). Because the $\omega_3$ and
%$\omega_4$ fields are generated from the vacuum, the dashed dotted
%lines in Fig.~1(d) are used to emphasize that there are no initial
%inputs. One underlying assumption is that the detectors D$_1$ and
%D$_2$ are not sensitive to the slight energy difference. This
%assumption is also a necessary requirement for the following
%beating experiment. In addition, Eq.~(\ref{eq:state}) is the
%essential result obtained in \cite{wen,du}.

The two-photon temporal coincidence counting rate $R_{c}$ between detectors D$_1$ and D$_2$ is
\begin{eqnarray}
R_{c}&=&\eta_1\eta_2\langle\Psi|E^{(-)}_1(\tau_1)E^{(-)}_2(\tau_2)E^{(+)}_2(\tau_2)E^{(+)}_1(\tau_1)|\Psi\rangle\nonumber\\
&=&|\langle0|E^{(+)}_2(\tau_2)E^{(+)}_1(\tau_1)|\Psi\rangle|^2=|A|^2,
\label{eq:definition}
\end{eqnarray}
which describes the second-order quantum coherence. In Eq.~(\ref{eq:definition}) $\eta_1$ and $\eta_2$ are the quantum detection efficiencies of two detectors, and $A$ is the two-photon amplitude given in Eq.~(\ref{eq:state}), respectively. Now we concentrate on the experimental setup shown in Fig.~1(a). Setting $\eta_1=\eta_2=100\%$ ideally and making use of Eq.~(\ref{eq:state}) give
\begin{eqnarray}
R_{c}=\frac{1}{4}\Big[e^{-2\gamma_g\tau}+e^{-2\gamma_2\tau}-2\cos(\Omega_e\tau-\phi)e^{-(\gamma_g+
\gamma_2)\tau}\Big].\label{eq:coin1}
\end{eqnarray}
The physics of Eq.~(\ref{eq:coin1}) is easy to be understood as follows: the first term corresponds to the correlation of biphotons from the central component in the Mollow triplet, the second term to the correlation of biphotons from two sidebands, and the third term to an interference between two previous correlations. Considering the beating experiment proposed here, an alternative explanation of Eq.~(\ref{eq:coin1}) can therefore be visualized as the first term gives the intensity of the biphotons following the short-short pathway in the Franson experiment, the second term gives the intensity of the biphotons following the long-long pathway, and the third term is the interference between the two biphotons. Equation~(\ref{eq:coin1}) shows that the pattern of the two-photon interference is a damped Rabi oscillation where the oscillation frequency is the effective Rabi frequency $\Omega_e$ and the damping rate is determined by the linewidths in the Mollow triplet. By subtracting accidental coincidences from the pump, the visibility in the center part of $R_c(\tau)$ can be changed between 0 and 1 by manipulating $\phi$. This change indicates a switch occurrence from photon anti-bunching-like effect to bunching-like effect. Alternatively, the interference occurring between two types of FWM processes can be switched from destructive to constructive, similar as the discussions presented in the case of a three-level double-$\Lambda$ system \cite{oh}. Equation~(\ref{eq:coin1}) can be also written as
\begin{eqnarray}
R_{c}=\frac{1}{2}\{\cosh[(\gamma_g-\gamma_2)\tau]-\cos(\Omega_e\tau-\phi)\}e^{-(\gamma_g+\gamma_2)\tau},\label{eq:coin2}
\end{eqnarray}
where $\cosh(x)=(e^x+e^{-x})/2$ is a hyperbolic function. If $\gamma_g\approx\gamma_2\approx\gamma$, Eq.~(\ref{eq:coin2}) reduces to a simple form
\begin{eqnarray}
R_c=\frac{1}{2}[1-\cos(\Omega_e\tau-\phi)]e^{-2\gamma\tau}.\label{eq:coin3}
\end{eqnarray}

Without the background noise, the two-photon temporal correlation is shown in Fig.~2. By adjusting the pump Rabi frequency, the atomic density, and the length of the medium, one can manipulate the center part in the two-photon interference (or beating) pattern. The (red) dotted line represents result using a weaker pump while the (blue) solid line represents a stronger pump case. The oscillation period using a weaker pump power is slower than that with a stronger pump power. The central parts (which are highlighted by a green rectangular) of the beating interference patterns are different: the (red) dotted curve almost goes to zero, indicating that the relative phase difference $\phi$ in this case is negligible; while the (blue) solid line is above zero, indicating that $\phi$ plays a major role in such a case. Hence, by changing the pump power, the range of the visibility in the center part is ranged from one to zero. As discussed in many papers \cite{kwait90,su,franson91}, 50$\%$ is the maximum visibility possible in a classical field approach to this sort of experiment.
%%%%%%%%%%%%%%%%%%%%%%%%%%%%%%%%%%%%%%%%%%%%%%%%%%
\begin{figure}[htb]
\centering\includegraphics[width=7cm]{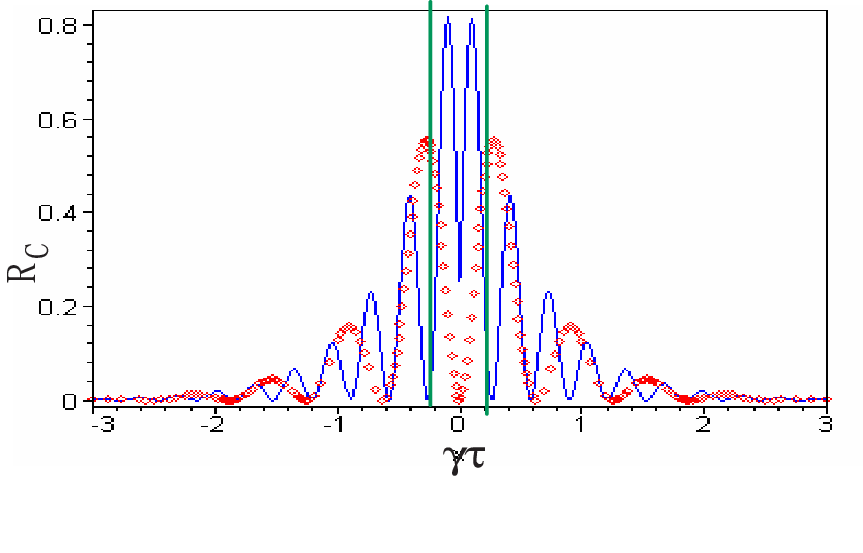}
\caption{(Color Online) Two-photon beating that is being observed with different pump powers.} \label{Fig.2}
\end{figure}
%%%%%%%%%%%%%%%%%%%%%%%%%%%%%%%%%%%%%%%%%%%%%%%%%%

Before proceeding the discussion, we note that a narrower filter would destroy the interference effect here. However,  since the Mollow triplet separation is tunable, this experiment may provide an interesting way of probing the quantum nature of the detection process. The interference will disappear when the separation of the Mollow
peaks approaches the fundamental timescales for photon absorption in the detector.

To realize this beating experiment, the natural bandwidth of biphotons from the phase matching conditions should be made larger than the linewidths of the Mollow triplet so that the intrinsic mechanism of biphoton generation is not erased by the phase matching. Experimentally, the optical depth of the medium should be small such that the two-photon wavepacket is determined by $\chi^{(3)}$ and not by the phase matching \cite{wen,wen2}. Realistically, two potential candidate configurations may be chosen as candidates for this type of experiment: (1) the system of cavity enhanced quantum dots, and (2) the laser cooled and trapped atomic ensemble. However, when operating in the resonant pumping case, (2) imposes a tougher challenge because atoms are often ``kicked out" of the trapping volume and this leads to the absorption dominant in the process. To avoid such problems, it may be advantageous to work with blue-detuned and red-detuned pumping arrangements.

\section{Conclusion}
In summary, a new beating experiment is proposed by using paired photons generated from a two-level system in the resonant pumping configuration. The Mollow triplet in the third-order nonlinear susceptibility indicates that two types of biphotons are created in the parametric interaction. The linear response to two types of biphotons allows tunable refractive indices with vanishing absorption to two sidebands in the Mollow triplet while as a unity index of refraction to the central component. The relative phase difference caused by different refractive indices is analogous to the interferometer proposed by Franson. By manipulating the pump power, the length of the material, and the atomic density, the visibility of the center part in the two-photon beating interference can be changed for 0 to 1 by subtracting background coincidences from the linear Rayleigh scattering of the pump beam. The observed interference pattern shows a switch from the destructive interference to the constructive in the FWM processes. Although the experiment is difficult, it is definitely possible to observe the nonzero effect in the center part of the beating pattern in the current labs. In addition, the experiment may be a way of probing the quantum nature of the detection process, since the interference will disappear when the separation of the Mollow peaks approaches the fundamental timescales for photon absorption in the detector.

\section*{Acknowledgments}
The authors are grateful to Yanhua Shih, James D. Franson, and Todd B. Pittman for fruitful discussions. We also thank the referees for insightful suggestions. J.-M. W. and M. H. R. were supported in part by the U.S. Army Research Office under MURI Grant W911NF-05-1-0197. S. D. acknowledges the start-up support from the Department of Physics at the Hong Kong University of Science and Technology E. O. was supported by the Office of Naval Research.

\end{document}